\documentstyle[aps,epsf,floats]{revtex}  

\begin{document}        

\baselineskip 14pt
\title{Recent Heavy-Flavor Measurements from OPAL}
\author{Thomas R. Junk}
\address{Carleton University\\ 1125 Colonel By Drive\\Ottawa, Canada, K1S 5B6}
%
\maketitle              

\begin{abstract}        
     A selection of recent heavy-flavor results from OPAL using the 
LEP1 data sample are presented.
The average polarization of b baryons in hadronic 
Z$^0$ decay has been measured
to be $-0.56^{+0.20}_{-0.13}{\mathrm (stat.)}\pm 0.09{\mathrm (syst.)}$ 
using semileptonic decays of $\Lambda_{\mathrm b}$
baryons.  A search has been conducted for 
the radially excited D$^{*\prime}$
and has produced a 95\% CL upper limit on its production of
$f({\mathrm Z}^0\rightarrow{\mathrm D}^{*\prime\pm}(2629))
\times {\mathrm Br}({\mathrm D}^{*\prime\pm}
\rightarrow {\mathrm D}^{*\pm}\pi^+\pi^-)
< 2.1\times 10^{-3}$.
Finally, the measurement of the product branching ratio
$f({\mathrm b}\rightarrow\Lambda_{\mathrm b})\times 
{\mathrm Br}(\Lambda_{\mathrm b}\rightarrow\Lambda {\mathrm X})=
(2.67\pm 0.38{\mathrm (stat)}
^{+0.67}_{-0.60}{\mathrm (syst.)})\%$ has been made.  
This measurement, along with
an earlier measurement of the product branching ratio
 $f({\mathrm b}\rightarrow\Lambda_{\mathrm b})\times 
{\mathrm Br}(\Lambda_{\mathrm b}\rightarrow\Lambda\ell {\mathrm X})$,
has been used to compute an updated $R_{\Lambda\ell} = 
{\mathrm Br}(\Lambda_{\mathrm b}\rightarrow\Lambda\ell {\mathrm X})/
{\mathrm Br}(\Lambda_{\mathrm b}\rightarrow\Lambda {\mathrm X})=
(8.0\pm 1.2{\mathrm (stat.)}\pm 0.9{\mathrm (syst.)})\%$,
consistent with the expected low semileptonic branching
fraction of the $\Lambda_{\mathrm b}$ inferred
from its short lifetime compared to the other b hadrons.
\end{abstract}          

\section{Measurement of the Average Polarization of b Baryons in Z$^0$ Decays}

In the Standard Model, b quarks produced in e$^+$e$^-$ collisions with
$\sqrt{s}\approx m_{\mathrm Z^0}$ have an average longitudinal
polarization of $-0.94$; the sign indicates that the b quarks are
predominantly left-handed.
The variation of this polarization with the angle of emission
of the b quark $\theta_{\mathrm b}$
\footnote{In the right-handed OPAL coordinate
system, the positive $z$ axis points along the e$^-$ beam and the $x$ axis points
towards the center for the LEP ring.  The polar and azimuthal angles are denoted by
$\theta$ and $\phi$, and the origin is
taken to be the center of the detector.} is only $\pm$2\%,
and the dependence on $\sin^2\!\theta_W$ is fairly weak: 
$\partial\langle P_{\mathrm b}\rangle/\partial\sin^2\theta_W \approx 0.63$.
This average polarization
of b quarks is the quantity $A_{\mathrm b}$ measured by SLD~\cite{SLDAB},
and is an ingredient
to the $A_{\mathrm FB}^0$ measurements from LEP~\cite{AFBLEP}.

In the heavy-quark limit, the spin of the b quark is expected to decouple from the
light degrees of freedom produced by the fragmentation processes.  The process of
stable hadron formation has a profound effect on the polarization retained in the
final state.  The pseudoscalar mesons B$^0$ and B$^\pm$ carry no polarization
information.
The vector meson B$^{*}$ decays by photon radiation which couples mainly to the light
degrees of freedom within the meson, and hence the radiated photons are very
nearly isotropic~\cite{FALKPESKIN}.  When directly produced, the $\Lambda_{\mathrm b}$
is expected to retain nearly the full polarization of the parent 
b quark~\cite{MANNELSCHULER}, where the light valence quarks form a spin-zero
system.  Some $\Lambda_{\mathrm b}$ mesons are produced from strong decays of
the $\Sigma_{\mathrm b}$
and the $\Sigma_{\mathrm b}^*$.  The energy difference between these states
is the hyperfine
splitting induced by the relative orientation of the b quark spin and the
spin-1 diquark.  These states are produced in a coherent superposition in the
hadronization process.  If the $\Sigma_{\mathrm b}$ and the  $\Sigma_{\mathrm b}^*$
  have narrower widths than their mass splittings,
then they decay incoherently; their lifetime is long enough to flip the spin
of the b quark.
If the states are broader than the splitting, then the b polarization retention
is higher.  The widths of these states are predicted to be of the order of the mass
splitting between them, hence substantial depolarization is expected for
these cascade decays~\cite{FALKPESKIN,KORNER96}.  A measurement of the average
polarization of $\Lambda_{\mathrm b}$ baryons tests both Heavy Quark
Effective Theory (HQET)
and also models of baryon formation.

The polarization of $\Lambda_{\mathrm b}$ baryons is most easily measured using their
semileptonic decays 
$\Lambda_{\mathrm b}\rightarrow\ell^-{\overline\nu}_\ell {\mathrm X}$~\footnote{Charge
conjugate processes are implied in this analysis and throughout this
article.}.  To conserve
angular momentum in the maximally parity-violating decay, the charged leptons $\ell^-$
are preferentially emitted antiparallel to the spin of the 
$\Lambda_{\mathrm b}$, while the
${\overline\nu}_\ell$'s are preferentially emitted parallel to the spin.  When boosted
into the laboratory frame, the charged lepton energy is on average higher than the
neutrino energy.  The spectra depend on 
the $\Lambda_{\mathrm b}$ momentum spectrum and the details of
the semileptonic decay, 
which are influenced by
$m_{\mathrm b}/m_{\mathrm c}$, QCD corrections, and form factors.  
The ratio of the average charged
lepton energy to the average neutrino energy partially cancels these effects and
is statistically more powerful than either energy alone~\cite{BONVICINIRANDALL}. 
This analysis is based on
the event-by-event distribution of the ratio $E_\nu/E_\ell$.
The ingredients needed for
the analysis are a purified sample of semileptonic $\Lambda_{\mathrm b}$ decays, 
an estimation of
the background contribution, and measurements of the lepton and neutrino energies.

The OPAL measurement is described in detail in Reference~\cite{OPALLBPOL} and is
summarized here.
The selection uses correlations between reconstructed 
$\Lambda\rightarrow {\mathrm p}\pi^-$
decays and high-$(p,p_t)$ leptons similar to the selection of~\cite{OPALLBLIFE},
but with improved electron identification, different lepton kinematic cuts, and the
inclusion of the requirement of a $\pi^+$, whose charge is required to be correlated
with the reconstructed $\Lambda$ to take advantage of the decay chain
$\Lambda_{\mathrm b}\rightarrow\Lambda_{\mathrm c}^+\rightarrow\Lambda\pi^+{\mathrm X}$.  
The $\pi^+$ is additionally required
to be consistent with coming from the common $\Lambda$-lepton vertex.

\begin{figure}
\centerline{\epsfxsize 3.0 truein \epsfbox{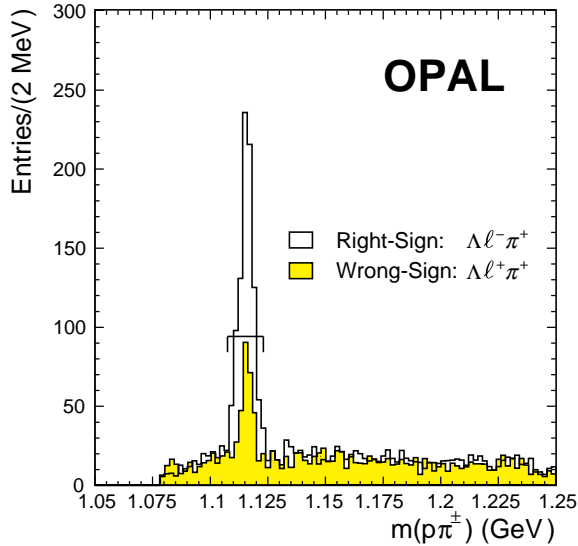}}   
\caption[]{Invariant mass distribution of p$\pi^-$ combinations in the
right-sign $\Lambda\ell^-\pi^+$ sample (open histogram) shown with the
distribution for wrong-sign $\Lambda\ell^+\pi^+$ combinations (shaded histogram);
the histograms are not added.  The signal region is shown in brackets.  The peak
in the wrong-sign distribution arises from genuine lambdas from fragmentation or
B meson decay paired with real or fake leptons.}
\label{PPIFIG}
\end{figure}

The invariant mass distribution of the selected p$\pi^-$ combinations is shown in 
Figure~\ref{PPIFIG}, separately for the $\Lambda\ell^-\pi^+$ (right-sign) and
$\Lambda\ell^+\pi^+$ (wrong-sign) combinations.  From a total of 4.3 million
hadronic Z$^0$ decays used in all three analyses presented in this article,
a total of 912 right-sign $\Lambda\ell\pi^+$ combinations
are selected with an overall b baryon purity of 69\%.  A total of 316 wrong-sign
combinations are also selected.  The wrong-sign combinations were used to estimate
the level of background as well as the shape of the background distribution of
$E_\nu/E_\ell$.  Part of the wrong-sign background consists of real leptons
from $\Lambda_{\mathrm b}$ decay combined with fragmentation $\Lambda$'s, 
which are preferentially
produced with the wrong sign because of conservation of baryon number; these events
contain the same polarization information as right-sign events.

To estimate the neutrino energy, the missing energy in the signal hemisphere
is computed as the difference between the expected total energy in the signal
hemisphere and the measured energy in the same hemisphere:
\begin{equation}
E_\nu=E_{beam} + \frac{m^2_{hemi}-m^2_{recoil}}{4E_{beam}} - E^{hemi}_{vis},
\end{equation}
where $E_{beam}$ is the beam energy, $E^{hemi}_{vis}$ is the measured visible energy
in the signal hemisphere, $m_{hemi}$ is the invariant mass of the observed
particles in the signal hemisphere and $m_{recoil}$ is the invariant mass of
the opposite hemisphere.  The estimated resolution of the neutrino energy
is 3.5~GeV.  The measurement error on the charged lepton momentum is negligible
by comparison, although the momentum scale and spectrum modeling introduce a
systematic uncertainty.
The distribution of the measured $E_\nu$ is compared between data
and Monte Carlo for events with a high-$(p,p_\perp)$ lepton and also for those
without, providing samples enriched in high-energy neutrinos and depleted in them.
Differences in the $E_\nu$ spectra are used to estimate corrections and systematic
errors.

\begin{figure}
\centerline{\epsfxsize 3.0 truein \epsfbox{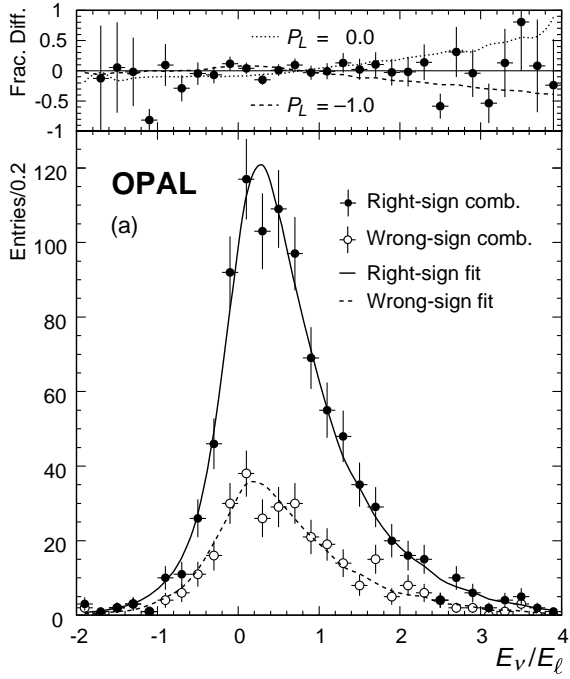}}   
\caption[]{
\small Bottom: distribution of the ratio of reconstructed energies $E_\nu/E_\ell$
for the right-sign combinations (solid circles) and wrong-sign combinations (open
circles).  The solid and dashed curves are the best-fit models to the distributions
in the right- and wrong-sign distributions, respectively.  Top: fractional residuals
in the fit to the right-sign sample, for the central value of $-56$\%, and also showing
models with $\langle P_L^{\Lambda_{\mathrm b}}\rangle$ values of $-1.0$ and 0.0.}
\label{PLFIT}
\end{figure}

The binned fit to the measured $E_\nu/E_\ell$ distribution was performed by comparing it
to fully simulated Monte Carlo samples generated with various values of 
the average polarization $\langle P_L^{\Lambda_{\mathrm b}}\rangle$.  
The right-sign and wrong-sign
spectra were fit simultaneously.  These distributions and the fit are shown in
Figure~\ref{PLFIT}.  The $\chi^2$ per degree of freedom is 0.91 and the fit value
is $\langle P_L^{\Lambda_{\mathrm b}}\rangle =-0.56^{+0.20}_{-0.13}$, where the
 errors are
statistical.

Table~\ref{LBPSYST} summarizes the sources of systematic uncertainty considered.
Added together in quadrature, the total systematic error is estimated to be
$\pm 0.09$.  The average longitudinal polarization of b baryons is measured to be
\begin{equation}
\langle P_L^{\Lambda_{\mathrm b}}\rangle=-0.56^{+0.20}_{-0.13}
{\mathrm (stat.)}\pm0.09{\mathrm(syst.)}.
\end{equation}
Including the effects of both statistical and systematic uncertainties, the 95\% 
confidence level (CL)
acceptance region is $-0.13\ge \langle P_L^{\Lambda_{\mathrm b}}\rangle \ge -0.87$, 
excluding for the first time zero polarization at the 95\% CL, and also disfavoring 
full preservation of the original polarization of $-0.94$ through the fragmentation
and hadronization processes.

\begin{table}
\begin{center}
\caption{Summary of systematic uncertainties in the 
measurement of $\langle P_L^{\Lambda_{\mathrm b}}\rangle$.}
\label{LBPSYST}
\begin{tabular}{lc|lc} 
Source of Uncertainty & $\Delta\langle P_L^{\Lambda_{\mathrm b}}\rangle$ &
Source of Uncertainty & $\Delta\langle P_L^{\Lambda_{\mathrm b}}\rangle$ \\
\tableline
$E_\nu$ resolution & $\pm 0.02$ & $\Lambda_{\mathrm c}$ polarization & $\pm 0.02$ \\ 
$E_\nu$ reconstruction & $\pm 0.05$ &  b$\rightarrow\tau$ & $\pm 0.01$ \\     
$E_\ell$ scale and shape & $\pm 0.03$ & Fitting method & $\pm 0.03$ \\          
Selection criteria & $\pm 0.02$ & Theoretical uncertainty (form factor & $\pm 0.03$ \\
Background fraction and shape & $\pm 0.04$ & 
modeling, QCD corrections, $m_{\mathrm c}/m_{\mathrm b}$ & \\      
b fragmentation & $\pm 0.03$ & & \\ \tableline
\multicolumn{4}{c}{\bf Total:\quad\quad $\pm 0.09$}  \\
\end{tabular}
\end{center}
\end{table}

\section{Search for the Radially Excited D$^{*\prime\pm}$(2629)}

In February 1998, the DELPHI collaboration reported 
evidence~\cite{DELPHIDSTARPRIME} for
a radially excited charm meson, labeled the D$^{*\prime}$.  The lowest-energy
radially excited state is called the D$^\prime$, but the D$^{*\prime\pm}$ is
easier to select because it is expected to decay to D$^{*\pm}\pi^+\pi^-$ with
a significant branching ratio.
DELPHI's measured signal rate is expressed in terms of the rates
of production of D$_1^0$ and D$_2^{*0}$ mesons in Z$^0$ decay:
\begin{equation}
R_{\mathrm DELPHI} = \frac{\langle N_{{\mathrm D}^{*\prime\pm}}\rangle
{\mathrm Br}({\mathrm D}^{*\prime +}\rightarrow
{\mathrm D}^{*+}\pi^+\pi^-)}
 {\sum_{J=1,2}  \langle N_{ {\mathrm D}_J^{(*)0} }\rangle 
{\mathrm Br}({\mathrm D}_J^{(*)0}\rightarrow {\mathrm D}^{*+}\pi^-)} =
0.48 \pm 0.18({\mathrm stat.})\pm 0.10({\mathrm syst.}).
\label{DELPHIDSTARPRIMEEQUATION}
\end{equation}
The 95\% CL upper limit on the width of the observed resonance is 15~MeV
and is limited by the detector resolution.  The observed mass is
2637$\pm$6~MeV.

The expected mass of the D$^{*\prime}$ is 2629$\pm$20~MeV~\cite{EXPECTMASSDSTARPRIME},
consistent with the DELPHI observation.  An estimate of the decay width
$\Gamma({\mathrm D}^{*\prime}\rightarrow{\mathrm D}^{*+}\pi^+\pi^-)$ using
a harmonic oscillator approximation~\cite{HARMONICOSCDSTARPRIME} is $<1$~MeV,
although a more recent computation by Melikhov and P\'ene~\cite{MELIKHOVPENE}
of the same partial width using heavy-quark
symmetry is between 120 and 160~MeV. Melikhov and P\'ene also estimate the 
width 
$\Gamma({\mathrm D}^{*\prime}\rightarrow{\mathrm D}^{(*)}\pi$ 
to be between 45 and 450~MeV, indicating some trouble in interpreting
the observation.  In addition, CLEO has searched for the D$^{*\prime +}$ in an
analysis sensitive only to 
e$^+$e$^-\rightarrow{\mathrm c{\overline c}}\rightarrow$D$^{*\prime +}$~\cite{SHIPSEY}
and has placed a limit of $R<0.16$ at 90\%~CL.

OPAL has searched for the D$^{*\prime +}$ and the selection procedure is
summarized here.  Charged D$^{*+}$ candidates
are selected in the decay mode D$^{*+}\rightarrow$D$^0\pi^+, $D$^0\rightarrow$K$^-\pi^+$.
The K$^-\pi^+$ combination is required to have its invariant mass within $\pm 2\sigma$ of 
the expected D$^0$ mass; the resolution is 25~MeV in both data and Monte Carlo.
The D$^{*+}$ is formed with a pion of opposite sign to the kaon and the scaled
energy of the D$^{*+}$, $x_E=E_{{\mathrm D}^{*+}}/E_{beam}$ is required to exceed 0.2.
The mass difference $\Delta m=m_{{\mathrm D}^{*+}}-m_{{\mathrm D}^0}$ must be between
142 and 149~MeV, corresponding to $\pm 3\sigma$ in the resolution.  Additional background
rejection is obtained by placing requirements on the helicity angle and the angle between
the kaon and the D$^0$ flight direction in the D$^0$ rest frame.  OPAL's
excellent d$E$/d$x$ capability~\cite{DEDX} is additionally used to purify the kaon sample.
The D$^{*\prime +}$ candidates are formed by combining the D$^{*+}$ candidates with
pairs of oppositely charged tracks, in order to explore DELPHI's observation in the
D$^{*\pm}\pi^+\pi^-$ decay mode.  These pions have loose d$E$/d$x$ requirements 
placed on them.  The reconstructed mass of the D$^{*\prime +}$ is corrected by
using the PDG value for the D$^{*+}$ mass and the measured difference between
the D$^{*\prime +}$ and the D$^{*+}$ candidates.

The analysis proceeds separately for events tagged as bottom and charm, in order
to suppress background from Z$^0\rightarrow {\mathrm u{\overline u},
d{\overline d}}$, and ${\mathrm s{\overline s}}$  events.
To select candidates from a sample enriched in 
Z$^0\rightarrow{\mathrm c{\overline c}}$ events,
the D$^{*+}$ candidate is required to have $x_E>0.4$, and the magnitude of
the vector sum of the momenta of the 
two pions added to form the D$^{*\prime +}$ candidate is required to be greater than
3.6~GeV.  To select candidates in Z$^0\rightarrow{\mathrm b{\overline b}}$ events,
the separation of the D$^0$ vertex from the primary divided by its error is
required to be larger than 4.0.  A loose requirement is made on the separation
between the D$^0$ vertex and the D$^{*\prime +}$ pion vertex to reduce the
combinatoric contamination.  Using a Monte Carlo simulation, the efficiency
for selecting D$^{*\prime +}$ mesons in Z$^0\rightarrow{\mathrm c{\overline c}}$ events
is estimated to be (7.1$\pm$0.6)\%, while for Z$^0\rightarrow{\mathrm b{\overline b}}$ it is
estimated to be (6.9$\pm$0.6)\%.  The combined signal expectation in OPAL is a factor of
1.1$\pm$0.1 times that of DELPHI, although the mass resolution in OPAL is
15~MeV and the background level is about 50\% higher, when the b and c analyses
are combined.

\begin{figure}
\centerline{\epsfxsize 3.0 truein \epsfbox{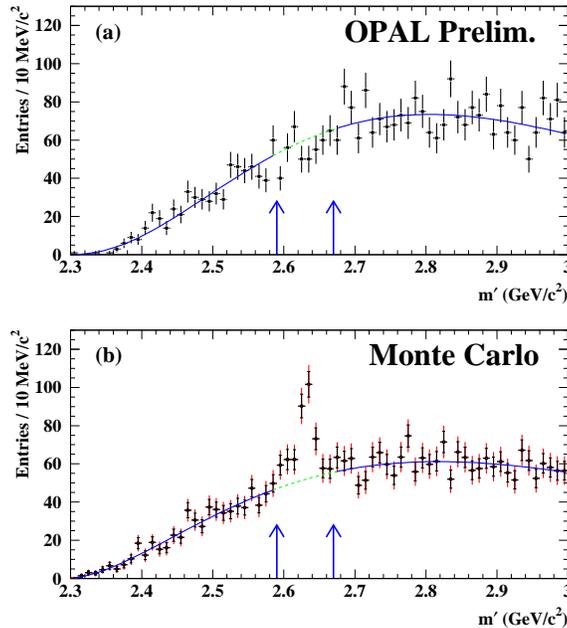}}   
\caption[]{
\small Reconstructed mass of D$^{*\prime\pm}$ candidates in OPAL data
(a), and in Monte Carlo simulation (b).  The D$^{*\pm}$ mass has
been constrained to the world average when reconstructing the
D$^{*\prime\pm}$ candidates (see text).  The normalization of the
signal in (b) is with respect to the DELPHI observation, and the
width is taken to be 0.0~GeV.  The solid error bars in (b) represent
the true Monte Carlo statistical error, while the thin bars correspond
to the errors of a sample the size of the data sample.  A smooth parameterization
of the background is shown for both the data and Monte Carlo, excluding the
expected signal region indicated with arrows.}
\label{OPALDSTARPRIMEFIG}
\end{figure}

The combined result of the bottom and charm analyses is shown in Figure~\ref{OPALDSTARPRIMEFIG},
where events that pass either of the two analyses enter the histogram once.  The
Monte Carlo expectation for a signal the size of DELPHI's observation is also
shown.  Limits have been obtained on the D$^{*\prime +}$ production rate by
defining a mass window between 2.59 and 2.67~GeV ($\pm 2\sigma$), and comparing
the observed candidate count against a background from a smooth parameterization
shown also in Figure~\ref{OPALDSTARPRIMEFIG}.  Systematic uncertainties, summarized
in Table~\ref{DSTARPRIMESYSTTABLE}, are incorporated in the limits.  The 95\% CL
upper limit on the number of D$^{*\prime +}$ candidates in the combined analysis
in the mass window is 32.8, with 443 events observed and 475 expected from the
background fit.  This can be converted into a limit on the production fraction:
\begin{equation}
f({\mathrm Z}^0\rightarrow{\mathrm D}^{*\prime\pm}(2629))\times
{\mathrm Br}({\mathrm D}^{*\prime +}\rightarrow{\mathrm D}^{*+}\pi^+\pi^-)
<2.1\times 10^{-3}\quad (95\%{\mathrm CL}).
\end{equation}
No signal is seen separately in the tagged b or c samples, and the upper limits
are
\begin{equation}
f({\mathrm c}\rightarrow{\mathrm D}^{*\prime +}(2629))\times
{\mathrm Br}({\mathrm D}^{*\prime +}\rightarrow{\mathrm D}^{*+}\pi^+\pi^-)
<1.2\times 10^{-2}\quad (95\%{\mathrm CL})
\end{equation}
and
\begin{equation}
f({\mathrm b}\rightarrow{\mathrm D}^{*\prime +}(2629))\times
{\mathrm Br}({\mathrm D}^{*\prime +}\rightarrow{\mathrm D}^{*+}\pi^+\pi^-)
<1.0\times 10^{-2}\quad (95\%{\mathrm CL}).
\end{equation}

The analysis has been checked by reconstructing the orbitally excited
D$^0_1$ and D$^{*0}_2$, following a very similar selection procedure,
except only one pion is added to the D$^{*+}$ candidate.  The rates
and widths of these are consistent with Monte Carlo expectations.  Combining
the limits above with the measurements of these rates, which share common
systematic uncertainties, OPAL sets a limit of $R<0.21$ at the 95\% CL, where
$R$ is defined in Equation~\ref{DELPHIDSTARPRIMEEQUATION}.

\begin{table}
  \caption{Summary of the systematic uncertainties on the selection efficiencies
  for D$^{*\prime +}$ in the c and b samples.}
  \label{DSTARPRIMESYSTTABLE}
  \begin{tabular}{lrrr}
  error source & \multicolumn{3}{c}{relative contribution} \\
               & c sample & b sample & combined sample \\
  \tableline
  \multicolumn{4}{l}{\it\quad\quad relative errors on Monte Carlo efficiency} \\
  Monte Carlo statistics           &  $9.3\%$ &  $9.1\%$ &  $6.2\%$ \\
  detector resolution              &  $8.9\%$ &  $0.7\%$ &  $5.1\%$ \\
  kaon d$E$d$x$                    &  $0.5\%$ &  $0.9\%$ &  $0.7\%$ \\
  pion d$E$d$x$                    &  $0.8\%$ &  $0.8\%$ &  $0.8\%$ \\
  D$^{*\prime +}$ width            &  $1.2\%$ & $16.7\%$ &  $3.6\%$ \\
  fragmentation modeling           &  $2.7\%$ &  $2.8\%$ &  $1.9\%$ \\
  production rates in b/c          &  none    &  none    &  $0.3\%$ \\
  \tableline
  total efficiency error           & $13.2\%$ & $19.3\%$ &  $9.1\%$ \\
  \tableline
  \tableline
  \multicolumn{4}{l}{\it\quad\quad relative errors on external branching ratios} \\
  branching ratio D$^{*+}\rightarrow$D$^0\pi^+$, D$^0\rightarrow$K$^-\pi^+$
                                   &  $3.1\%$ &  $3.1\%$ &  $3.1\%$ \\
  error on $\Gamma_{\mathrm q{\overline q}}/\Gamma_{\mathrm had}$     
                                   &  $4.5\%$ &  $0.6\%$ &  none    \\
  \tableline
  total branching ratio error     &  $5.5\%$ &  $3.1\%$ &  $3.1\%$ \\
  \end{tabular}
\end{table}

\section{Measurement of $\lowercase{f}({\mathrm \lowercase{b}}\rightarrow
\Lambda_{\mathrm \lowercase{b}})\times 
{\mathrm B\lowercase{r}}(\Lambda_{\mathrm \lowercase{b}}\rightarrow\Lambda {\mathrm X})$}

The measurement of $f({\mathrm b}\rightarrow\Lambda_{\mathrm b})\times 
{\mathrm Br}(\Lambda_{\mathrm b}\rightarrow\Lambda {\mathrm X})$ has
been carried out mainly because of its interest in interpreting measurements of
$f({\mathrm b}\rightarrow\Lambda_{\mathrm b})\times 
{\mathrm Br}(\Lambda_{\mathrm b}\rightarrow\Lambda\ell^-{\overline \nu} {\mathrm X})$
to extract the quantity 
$R_{\Lambda\ell}={\mathrm Br}(\Lambda_{\mathrm b}
\rightarrow\Lambda\ell^-{\overline \nu} {\mathrm X})/
{\mathrm Br}(\Lambda_{\mathrm b}\rightarrow\Lambda {\mathrm X})$, which is 
an approximate measure of
semileptonic branching fraction of the $\Lambda_{\mathrm b}$.  
The average lifetime of the
$\Lambda_{\mathrm b}$ has been measured to be approximately 80\% 
that of the B~mesons, while
theoretical predictions can accommodate ratios 
between 90\% and 100\%~\cite{ABBANEODPF}.
It is therefore important to check
the shortened lifetime with a measurement of the semileptonic branching ratio,
which is expected to be proportional to the lifetime within the B hadron family,
provided that the leptonic
decay currents are the same for the b~baryons and the B~mesons.
A second reason for measuring
this product branching ratio is that it allows one to extract 
${\mathrm Br}(\Lambda_{\mathrm b}\rightarrow\Lambda {\mathrm X})$.

This analysis is fully described in Reference~\cite{OPALPRODBR} and a summary is provided
here.  The procedure is to identify $\Lambda\rightarrow$p$\pi^-$ decays in b-tagged
events, and then to use the momentum, $p$, and transverse momentum, $p_t$
of each $\Lambda$ with respect to its jet, to separate the contributions from 
b~baryon decay, B~meson decay, and $\Lambda$'s produced by the fragmentation process.

Events are tagged as Z$^0\rightarrow{\mathrm b{\overline b}}$ if at least one jet
tags as a b using displaced secondary vertices as described in~\cite{OPALBTAG}.
The estimated overall efficiency is about 21\% per jet and the purity is about 
96\% per event.
The $\Lambda$ selection requires that the reconstructed mass 
of the p$\pi^-$ combination be
within 8~MeV of $m_\Lambda$, that it be farther than 6~MeV from the K$^0$ mass when
interpreted as a $\pi^+\pi^-$ pair, that the flight distance be at least 8.0~cm
from the interaction point when projected into the $xy$ plane, that the momentum points
back at the interaction point, that it have a momentum exceeding 5~GeV, and that it 
be within 0.2 radians of its jet axis.  In addition, OPAL's d$E$/d$x$ capability is
used to require that the proton's candidate track's ionization is consistent with
that expected from a proton.

\begin{figure}
\centerline{\epsfxsize 4.0 truein \epsfbox{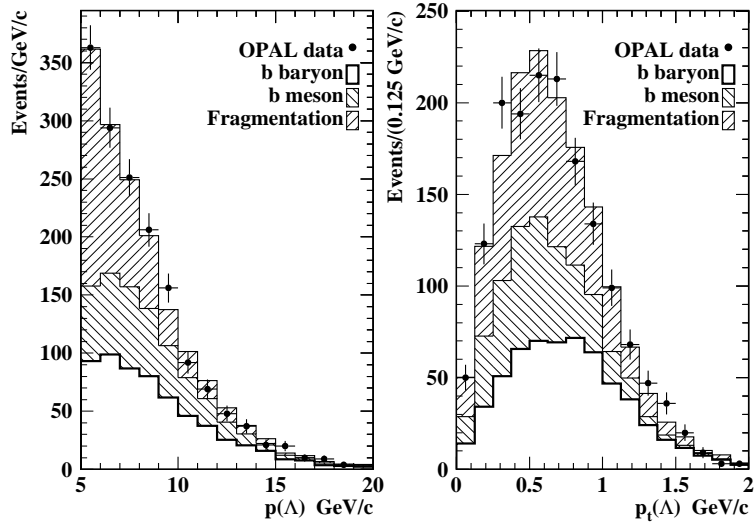}}   
\caption[]{Distributions of $\Lambda$ $p$ and $p_t$ in data (filled circles
with error bars) and Monte Carlo (histogram).  The three sources, fragmentation,
B~meson decay, and $\Lambda_{\mathrm b}$ decay are normalized to their
fitted fractions.}
\label{LSFITFIG}
\end{figure}

The Monte Carlo is used to provide distributions of $p$ and $p_t$~
\footnote{The $p_t$ of the $\Lambda$ is measured with respect to the nearest jet
axis, the calculation of which includes the $\Lambda$ momentum.}
for $\Lambda$'s
from each of the three sources, b~baryon decay, B~meson decay, and fragmentation.
Other backgrounds contribute 3\% of the sample, dominated by D$^+$ mesons tagging
as B~mesons because of their long lifetime.  Because the D$^+$ does not have
enough mass to decay into a $\Lambda$ and another baryon, the main contribution comes
from fragmentation $\Lambda$'s in events with tagged D$^+$ mesons.  These are incorporated
into the fragmentation lambda portion of the fit.  These distributions of $p$ and $p_t$
are jointly fit for the fractions of $\Lambda$'s in each of the three categories, and
the results are shown in Figure~\ref{LSFITFIG} and summarized in Table~\ref{LAMBDABRTAB1}.
The momentum is a powerful discriminator between b~baryon decay and fragmentation,
while the transverse momentum is effective for separating b~baryon decays from B~meson
decays, although this separation is more difficult, which is reflected in the correlations
shown in Table~\ref{LAMBDABRTAB1}.  A summary of the systematic errors on the fit
fractions is given in Table~\ref{LAMBDABRTAB2}.  Checks of systematic biases in the
$p$ and $p_t$ spectra were made by comparing events with $\Lambda$-lepton pairs
and comparing the averages of $p$ and $p_t$ in the right- and wrong-sign samples
to obtain estimations for a pure sample of semileptonic $\Lambda$ decays.

\begin{table}
\begin{center}
\caption{Results of the fit for the fractional contributions to the $\Lambda$ sample
  in b-tagged events.  The correlations between the fit fractions are also shown.}
\label{LAMBDABRTAB1}
\begin{tabular}{lcccc}
Source & Fit fraction (\%) & \multicolumn{3}{c}{Correlations}\\
 & & $\Lambda_{\mathrm b}\rightarrow\Lambda {\mathrm X}$ & 
frag$\rightarrow\Lambda {\mathrm X}$ & B$\rightarrow\Lambda {\mathrm X}$ \\\tableline
$\Lambda_{\mathrm b}\rightarrow\Lambda {\mathrm X}$ & 
37.4 $\pm$ 5.3 & 1.0   & --0.15 & --0.64 \\
frag$\rightarrow\Lambda {\mathrm X}$ & 37.1 $\pm$ 5.1 & --0.15 & 1.0   & --0.61 \\
B$\rightarrow\Lambda {\mathrm X}$ & 25.5 $\pm$ 6.6    & --0.64 & --0.61 & 1.0 \\
\end{tabular}
\end{center}
\end{table}

\begin{table}
\begin{center}
\caption{Systematic uncertainties on the 
$\Lambda_{\mathrm b}$ fit fraction, $f_{\Lambda_{\mathrm b}}$.}
\label{LAMBDABRTAB2}
\begin{tabular}{lcc}
Sources of Systematic Errors for $f_{\Lambda_{\mathrm b}}$ & negative & positive \\
 & errors & errors \\\tableline
$p(\Lambda$) and $p_t(\Lambda)$ from B$\rightarrow\Lambda {\mathrm X}$ &
 $-3.5$\% & 3.5\% \\
$p(\Lambda)$ of Fragmentation $\Lambda$'s & $-1.3$\% & 0.8\% \\
$p_t(\Lambda)$ of Fragmentation & $-12.5$\% & 12.7\% \\
$p(\Lambda)$ from $\Lambda_{\mathrm b}\rightarrow\Lambda {\mathrm X}$ &
 $-2.9$\% & 4.8\% \\
$p_t(\Lambda)$ from $\Lambda_{\mathrm b}\rightarrow\Lambda {\mathrm X}$ & $-16.6$\% & 19.8\% \\
Tracking Uncertainty & $-2.7$\% & 2.7\% \\\tableline
Total                                & $-21.5$\%   & 24.4\% \\
\end{tabular}
\end{center}
\end{table}

The other important ingredient in the analysis is to understand the efficiency of the
b-tag for jets containing a $\Lambda_{\mathrm b}$.  This can be accomplished by counting
events with one and two b-tagged jets while assuming the value of $R_{\mathrm b}$ and
the efficiency to tag non-b${\mathrm\overline b}$ events and solving for $\epsilon_{\mathrm b}$,
the b-tagging efficiency, using
\begin{equation}
  R_{\mathrm b}\epsilon_{\mathrm b} + 
(1-R_{\mathrm b})\epsilon_{udsc} = f_{\mathrm 1v},\quad{\mathrm and}
\end{equation}
\begin{equation}
  R_{\mathrm b}\epsilon_{\mathrm b}^2 + 
(1-R_{\mathrm b})\epsilon_{udsc}^2 = f_{{\mathrm 2v}}.
\end{equation}
Large biases enter, however, from the facts that 1) 
the $\Lambda_{\mathrm b}$ has a shorter
lifetime than the B~mesons and therefore tags less often, 2) the $\Lambda$ decay
products are not included in the secondary vertex, leaving a lower average multiplicity
and momentum of charged tracks which do participate in the secondary vertex, and
3) that if a high-momentum fragmentation $\Lambda$ is present, the b~hadron will
have less momentum than average.  These biases are accounted for by counting
the number of selected $\Lambda$'s in hemispheres with b-tags and hemispheres opposite
b-tags.  The tag efficiency corrections for each class of $\Lambda$ are modeled in Monte
Carlo, and the overall efficiency correction is scaled to match that of the data, obtained
from the $\Lambda$-tag statistics.  The scaled correction 
for the $\Lambda_{\mathrm b}\rightarrow\Lambda {\mathrm X}$
component is then used as the correction, which amounts a factor of $0.49\pm0.06$,
for an estimated event b-tag efficiency of $(29.4\pm 1.5)\%$.

The product branching ratio is then formed as follows:
\begin{equation}
f({\mathrm b}\rightarrow\Lambda_{\mathrm b})\times 
{\mathrm Br}(\Lambda_{\mathrm b}\rightarrow\Lambda {\mathrm X}) =
\frac{N_{\Lambda}f_{\Lambda_{\mathrm b}}}{2R_{\mathrm b} 
N_{mh}\epsilon_\Lambda\epsilon^b_{sig}
{\mathrm Br}(\Lambda\rightarrow{\mathrm p}\pi^-)},
\end{equation}
where $N_\Lambda$ is the number of $\Lambda$ candidates in the final sample,
$N_{mh}$ is the number of hadronic Z$^0$ decays in the final sample, $\epsilon^b_{sig}$
is the b-tagging efficiency for events with $\Lambda_{\mathrm b}$'s in them.
Table~\ref{LAMBDABRTAB3} gives a summary of the uncertainties on the measurement
of the product branching ratio.
Combining these factors with the appropriate systematic errors added in quadrature yields
\begin{equation}
f({\mathrm b}\rightarrow\Lambda_{\mathrm b})\times 
{\mathrm Br}(\Lambda_{\mathrm b}\rightarrow\Lambda {\mathrm X}) =
(2.67\pm 0.38({\mathrm stat.})^{+0.67}_{-0.60}({\mathrm syst.}))\%.
\end{equation}
Combining this measurement with a previous measurement of the product branching ratio
using a companion baryon technique~\cite{COMPANIONBARYON} yields a more precise
estimation, as the overlap in the selected event sample is less than 20\%.  The
combined measurement is
\begin{equation}
f({\mathrm b}\rightarrow\Lambda_{\mathrm b})\times 
{\mathrm Br}(\Lambda_{\mathrm b}\rightarrow\Lambda {\mathrm X}) =
(3.50\pm 0.32({\mathrm stat.})\pm 0.35({\mathrm syst.}))\%,
\end{equation}
in agreement with a measurement from DELPHI~\cite{DELPHIPRODUCTBR} 
of $(2.2^{+1.3}_{-0.8})\%$.
Using the measured value of $f({\mathrm b}\rightarrow\Lambda_{\mathrm b})$ in hadronic
Z$^0$ decays of $(10.1^{+3.9}_{-3.1})\%$~\cite{CASO}, one obtains
\begin{equation}
{\mathrm Br}(\Lambda_{\mathrm b}\rightarrow\Lambda {\mathrm X}) = (35^{+14}_{-12})\%.
\end{equation}

\begin{table}
\begin{center}
\caption{Quantities needed for evaluating the
  product branching ratio, and their impact on the uncertainty of the
  measurement.}
\label{LAMBDABRTAB3}
\begin{tabular}{lcc}
\hspace{5mm} & & Uncertainty on \\
\hspace{5mm} & & $f({\mathrm b}\rightarrow\Lambda_{\mathrm b})\times
  {\mathrm Br}(\Lambda_{\mathrm b}\rightarrow\Lambda {\mathrm X})$ \\\tableline
\hspace{5mm}$N_{\Lambda}$  & $1582\pm40({\mathrm stat.})$  & 
$\pm 0.07({\mathrm stat.})$ \\
\hspace{5mm}$f_{\Lambda_{\mathrm b}}$    & $0.374\pm
0.053({\mathrm stat.})^{+0.091}_{-0.080}({\mathrm syst.})$ & 
$0.38({\mathrm stat.})^{+0.65}_{-0.57}({\mathrm syst.})$ \\
\hspace{5mm}$R_{\mathrm b}$          & $0.2169\pm 0.0012({\mathrm syst.})$  & 
$\pm 0.01({\mathrm syst.})$ \\ 
\hspace{5mm}$N_{mh}$       & 2323302 & --- \\
\hspace{5mm}$\epsilon_\Lambda$ & $0.117\pm 0.006({\mathrm syst.})$ & 
$\pm 0.14({\mathrm syst.})$ \\
\hspace{5mm}$\epsilon_{sig}^b$ & $0.294\pm 0.015({\mathrm syst.})$ &  
$\pm 0.14({\mathrm syst.})$ \\
\hspace{5mm}${\mathrm Br}(\Lambda\rightarrow{\mathrm p}\pi^-$
     & $0.639\pm 0.005({\mathrm syst.})$ &  $\pm 0.02({\mathrm syst.})$ \\
\end{tabular}
\end{center}
\end{table}

\section{Summary}

OPAL has produced a wide spectrum of heavy-flavor physics results in the past year.
In addition to the measurements described here of the b~baryon longitudinal polarization,
the search for the radial excitation D$^{*\prime +}$, and the measurement of the
product branching ratio 
$f({\mathrm b}\rightarrow\Lambda_{\mathrm b})\times 
{\mathrm Br}(\Lambda_{\mathrm b}\rightarrow\Lambda {\mathrm X})$,
OPAL has completed its final $R_{\mathrm b}$ measurement~\cite{OPALRB98}, 
studied CP violation in the
$J/\Psi$-K$^0_s$ channel~\cite{SIN2BETA}, measured the $B^{\pm}$ and B$^0$ lifetimes 
using a topological
tag~\cite{TOPOLOGICALTAG}, and measured the semileptonic branching fraction
of inclusive b hadrons~\cite{SEMILEPTONIC}.  Each analysis
uses OPAL's full LEP1 data sample and are among the most precise published
measurements of their respective quantities.

\end{document}